\documentclass[aps,showpacs]{revtex4-2}
\usepackage{amsmath,amsfonts,bm}
\usepackage[dvips]{graphicx}
\usepackage[all]{xy}
\usepackage{color}
\def\:{\colon\!}

\begin{document}

\title{Heisenberg-Uncertainty of Spatially-Gated Electromagnetic Fields}

\author{Vladimir Y. Chernyak$^{a, b}$, and Shaul Mukamel$^{c, d}$}

\affiliation{$^a$Department of Chemistry, Wayne State University, 5101 Cass Ave, Detroit, Michigan 48202, USA}
\affiliation{$^b$Department of Mathematics, Wayne State University, 656 W. Kirby, Detroit, Michigan 48202, USA}
\affiliation{$^c$Departmnet of Chemistry, University of California, Irvine, CA 92614, USA}
\affiliation{$^d$Departmnet of Physics and Astronomy, University of California, Irvine, CA 92614, USA}
\email{chernyak@chem.wayne.edu, smukamel@uci.edu}


\date{\today}

\begin{abstract}
A Heisenberg uncertainty relation is derived for spatially-gated electric $\Delta E$ and magnetic $\Delta H$ fields fluctuations. The uncertainty increases for small gating sizes which implies that in confined spaces the quantum nature of the electromagnetic field must be taken into account. Optimizing the state of light to minimize $\Delta E$ at the expense of $\Delta H$, and vice versa should be possible. Spatial confinements and quantum fields may alternatively be realized without gating by interaction of the field with a nanostructure. Possible applications include nonlinear spectroscopy of nanostructures and optical cavities and chiral signals.
\end{abstract}


\maketitle

\section{Introduction}
\label{sec:inro}

The electric and the magnetic field operators $\hat{\bm{E}}(\bm{r})$ and $\hat{\bm{H}}(\bm{r}')$  at two different points in space do not commute. This implies the existence of a Heisenberg uncertainty relation between them. In his 1927 Chicago lecture notes \cite{Heisenberg}, Heisenberg had calculated this uncertainty for fields averaged over a box of size $ l$ and obtained $(\Delta E) (\Delta H) > \hbar c/l^{4}$. This implies that for a sufficiently small box, electromagnetic field fluctuations are strong enough so that the quantum nature of the fields may not be ignored. Heisenberg did not have a particular application in mind but was rather interested in clarifying a fundamental issue: the corpuscular vs wave nature of photons. Thanks to recent advances in nano optics \cite{Novotny-06} this uncertainty may be tested experimentally in nanostructures. Here we examine it for spatially-gated \cite{Novotny-06} electric and magnetic fields. An important consequence of our study is that spectroscopy of nanostructures may not be fully described by classical fields since this uncertainty may not be neglected. Spatially-gated fields thus have an intrinsic quantum nature that should have experimental signatures.

To derive the Heisenberg uncertainty relation for spatially gated electric and magnetic fields, we introduce two vector gate functions $\eta$ and $\gamma$, associated with the electric and magnetic fields that give rise to two gate function dependent gauge invariant Hermitian gated electric and magnetic field operators, 
defined by
\begin{eqnarray}
\label{E-H-gated-operators} \hat{E}(\eta) = \int d\bm{r} \bm{\eta}(\bm{r}) \cdot \hat{\bm{E}}(\bm{r}), \;\;\; \hat{H}(\gamma) = \int d\bm{r} \bm{\gamma}(\bm{r}) \cdot \hat{\bm{H}}(\bm{r}).
\end{eqnarray}
In the following sections we shall derive the commutation relation and the corresponding Heisenberg uncertainty relation that have the form
\begin{eqnarray}
\label{commute-uncertain} [\hat{E}(\eta),\hat{H}(\gamma)] = -i\hbar h(\eta, \gamma) \hat{I}, \;\;\; (\Delta \hat{E}(\eta)) (\Delta \hat{H}(\gamma)) \ge (\hbar/2) |h(\eta, \gamma)|.
\end{eqnarray}
In Eq.~(\ref{commute-uncertain}) we have used a standard notation $(\Delta \hat{Q}) = (\Delta \hat{Q})_{\Psi}$ for the uncertainty of a Hermitian operator acting in Hilbert space (and thus defining an observable) in a quantum state $|\Psi\rangle$,
\begin{eqnarray}
\label{define-uncertainty} (\Delta \hat{Q})_{\Psi} = \sqrt{\langle \Psi|(\hat{Q} - \langle \Psi|\hat{Q}|\Psi \rangle \hat{I})^{2}|\Psi \rangle}.
\end{eqnarray}
Details of the derivation of Eq.~(\ref{commute-uncertain}), based on standard theory of light-matter interactions~\cite{Mandel-Wolf,Andrews-18}, are presented in appendices~\ref{sec:EM-quant-uncert} and~\ref{sec:EM-uncertain-coherent-details}.

The connection between the uncertainty relation and commutator depends on the particle statistics and holds for any system of bosons. It applies in our case, since the EM field is a system of free bosons (photons). By applying a proper quantization procedure we can compute the commutator in Eq.~(\ref{commute-uncertain}), resulting in
\begin{eqnarray}
\label{expression-for-c}  h(\eta, \gamma) = 4 \pi c \int d\bm{r} ({\rm curl} \,\bm{\eta}(\bm{r})) \cdot \bm{\gamma}(\bm{r}) = - 4 \pi c \int d\bm{r} ({\rm curl} \,\bm{\gamma}(\bm{r})) \cdot \bm{\eta}(\bm{r}).
\end{eqnarray}
In deriving Eq.~(\ref{expression-for-c}) one should note the fact that the electromagnetic field is a system with first class constraints (using Dirac's terminology \cite{Dirac}) that give rise to gauge invariance, or, stated differently, that photons have transverse polarization.

\section{Commutation Rules and the Heisenberg Uncertainty of Gaussian Gated Fields}
\label{sec:EM-quant-uncert-G-gate}

The contribution to the measurement of the electric or the magnetic field at point $\bm{r}$ in space comes from some small region around $\bm{r}$, due to the spatial error bar of a measurement device. Here we is describe this by using Gaussian gates for both electric and magnetic fields. To that end we introduce a family of Gaussian approximations for the Dirac $\delta$-function
\begin{eqnarray}
\label{delta-Gauss} \delta (\bm{r}; l) = \frac{1}{(\sqrt{2\pi} l)^{3}} \exp(-(\bm{r}^{2}/2l)^{2}),
\end{eqnarray}
and the corresponding Gaussian-gated electric and magnetic fields
\begin{eqnarray}
\label{E-H-Gauss-gated} \bm{E} (\bm{r}; l) = \int d\bm{r}' \delta (\bm{r}-\bm{r}'; l) \bm{E}(\bm{r}'), \;\;\; \bm{H} (\bm{r}; l) = \int d\bm{r}' \delta (\bm{r}-\bm{r}'; l) \bm{H}(\bm{r}').
\end{eqnarray}
The Heisenberg uncertainty relation for $(\Delta E_{j} (\bm{r}; l)) (\Delta H_{s} (\bm{r}'; l))$ is completely determined by the commutator of the corresponding gated operators \cite{Mandel-Wolf,Salam}
\begin{eqnarray}
\label{E-H-G-gated-commute} && [\hat{E}_{j}(\bm{r}; l), \hat{H}_{s}(\bm{r}'; l)] = 4\pi c i\hbar \varepsilon_{jsm} \frac{\partial}{\partial r_{m}} \delta (\bm{r} - \bm{r}'; \sqrt{2}l), \nonumber \\ && \frac{\partial}{\partial r_{m}} \delta (\bm{r} - \bm{r}'; \sqrt{2}l) = - \frac{1}{\sqrt{2}l} \frac{r_{m} - r_{m}'}{\sqrt{2}l} \frac{1}{(\sqrt{4\pi}l)^{3}}  \exp \left(- \frac{(\bm{r} - \bm{r}')^{2}}{4l^{2}}\right),
\end{eqnarray}
these immediately lead to the Heisenberg uncertainty relation
\begin{eqnarray}
\label{E-H-G-gated-Heisenberg} (\Delta E_{j} (\bm{r}; l)) (\Delta H_{s} (\bm{r}'; l)) \ge 4\pi \frac{\hbar c}{\sqrt{2}l} \frac{1}{(\sqrt{4\pi}l)^{3}} \frac{|r_{m} - r_{m}'|}{\sqrt{2}l} \exp \left(- \frac{(\bm{r} - \bm{r}')^{2}}{4l^{2}}\right) |\varepsilon_{jsm}|.
\end{eqnarray}
To derive Eq.~(\ref{E-H-G-gated-commute}) we applied Eq.~(\ref{EM-P-bracket-k-6}) and computed $h (\eta, \gamma)$ for Gaussian gate functions making use of Eq.~(\ref{expression-for-c}). The Gaussian integrations are easily performed. A similar calculation shows that in the case of not necessarily identical Gaussian gates, the Heisenberg uncertainty relation for $(\Delta E_{j} (\bm{r}; l)) (\Delta H_{s} (\bm{r}'; l'))$ still has a form of Eq.~(\ref{E-H-G-gated-Heisenberg}), with the gate size $l$ in the r.h.s. replaced by
\begin{eqnarray}
\label{E-H-G-gated-Heisenberg-2} l_{\rm eff} = \sqrt{\frac{l^{2} + (l')^{2}}{2}}.
\end{eqnarray}

Eq.~(\ref{E-H-G-gated-Heisenberg}) provides a clear physical interpretation for the $1/l$ parameter in the Heisenberg's uncertainty formula. Eq.~(\ref{E-H-G-gated-Heisenberg}) depends on a single parameter: the gate size $l$. The $(\hbar c)/(\sqrt{2}l)$ factor represents the $1/l$ Heisenberg's parameter in energy units; the $\sqrt{2}$ factor is specific to the Gaussian form of the gate. More precisely the $(\hbar c)/(\sqrt{2}l)$ factor is the energy of a photon, whose wavelength is given by the gate size. The second factor $\frac{1}{(\sqrt{4\pi}l)^{3}}$ is of the order of the inverse volume of the gate region. Their product thus represents the energy density of a photon, restricted to the gate region (according to the Heisenberg uncertainty principle for a photon, considered as a quantum particle). It is worth noting that $(\Delta E)(\Delta H)$ has units of energy density. The dimensionless product of the last two factors in Eq.~(\ref{E-H-G-gated-Heisenberg}) characterizes the overlap of the gates, so that when $|\bm{r} - \bm{r}'| \gg l$ the uncertainty vanishes. It is also interesting to note that at $\bm{r} = \bm{r}'$ the uncertainty vanishes as well, provided the gate profiles are identical. One way to look at this dependence is that even if the original field is classical and contains many photons, the nano-gated field may contain only a few photons, and for a sufficiently short gate it will, therefore, show quantum effects.

To get a sense of the fluctuation magnitudes, we compare $(\Delta E) (\Delta H)$ for a Gaussian gate with $l = 1 \, {\rm nm} = 10^{-9} \, {\rm m}$ with the $EH$ product in a pulse with the intensity $I = 10^{15} \, {J} \cdot {\rm s}^{-1} \cdot {\rm cm}^{-2} = 10^{19} \, {J} \cdot {\rm s}^{-1} \cdot {\rm m}^{-2}$. The maximal value of the dimensionless function (the product of the last two factors in the r.h.s. of Eq.~(\ref{E-H-G-gated-Heisenberg})) is $1/\sqrt{e} \approx 0.6$, so that we have for $l = 1 \, {\rm nm} = 10^{-9} \, {\rm m}$ and $hc \approx 2 \times 10^{-25} \, {\rm J} \cdot {\rm m}$,
\begin{eqnarray}
\label{E-H-Heisenberg-example} (\Delta E) \cdot (\Delta H) \sim 4\pi \frac{\hbar c}{\sqrt{2}l} \frac{1}{(\sqrt{4\pi}l)^{3}} \frac{1}{\sqrt{e}} = \frac{1}{4\pi \sqrt{2\pi} \sqrt{e}} \frac{hc}{l^{4}} \approx \frac{1}{4\pi \sqrt{2\pi} \sqrt{e}} \frac{2 \times 10^{-25} \, {\rm J} \cdot {\rm m}}{10^{-36} {\rm m}^{4}} \approx \frac{1}{4\pi \sqrt{2\pi} \sqrt{e}} 2 \times 10^{11} {\rm J} \cdot {\rm m}^{3}.
\end{eqnarray}
Making use of the Poynting vector
\begin{eqnarray}
\label{Poyn-vector} \bm{S} = - \frac{c}{4\pi} [\bm{E}, \bm{H}], \;\;\;\;\;\; {\rm so} \; {\rm that} \;\; I = S = \frac{c}{4\pi} EH
\end{eqnarray}
we have
\begin{eqnarray}
\label{EH-example} E \cdot H = \frac{4\pi}{c} I \approx 4\pi \frac{10^{19} \, {\rm J} \cdot {\rm s}^{-1} \cdot {\rm m}^{-2}}{3 \times 10^{8} \, {\rm m} \cdot {\rm s}^{-1}} \approx 4\pi \frac{1}{3} \times 10^{11} \, {\rm J} \cdot {\rm m}^{-3},
\end{eqnarray}
so that
\begin{eqnarray}
\label{Delta-to-EH-example} \frac{(\Delta E) \cdot (\Delta H)}{E \cdot H} \approx \frac{1}{(4\pi)^{2} \sqrt{2\pi} \sqrt{e}} \cdot 6 \approx \frac{0.6}{100} \cdot 6 \approx 4 \times 10^{-2}.
\end{eqnarray}
Finally setting $E = H$ for a plane wave and $\Delta E = \Delta H$ for a coherent state of the electromagnetic field we obtain
\begin{eqnarray}
\label{Delta-to-EH-example-2} \frac{\Delta E}{E} \approx 0.2
\end{eqnarray}
implying that for a nanometer gate quantum effects are non-negligible even for a very strong pulse, which is usually thought of as classical.

\section{Uncertainties of the Gated Electric and Magnetic Fields for a Coherent State}
\label{sec:EM-uncertain-coherent}

We now compute the uncertainties of the gated and electric and magnetic fields prepared in a coherent sate. Since, in the coordinate representation, this state is simply a displaced ground state, the problem is reduced to computing the variance of the quantum fluctuations of the gated fields in vacuum, i.e., we have
\begin{eqnarray}
\label{uncertain-EM-coherent} (\Delta E(\eta))^{2} = \langle \Omega| (\hat{E} (\eta))^{2} |\Omega \rangle, \;\;\; (\Delta H(\gamma))^{2} = \langle \Omega| (\hat{H} (\gamma))^{2} |\Omega \rangle.
\end{eqnarray}
The computation is carried out by introducing the photon creation/annihilation operators, see appendix~\ref{sec:EM-uncertain-coherent-details} for details.

Applying the continuum limit to Eqs.~(\ref{compute-Delta-E-2}) and (\ref{Delta-H}) we obtain
\begin{eqnarray}
\label{Delta-E-H-continuous} (\Delta E(\eta))^{2} &=& \int \frac{d\bm{k}}{(2\pi)^{3}} 2\pi\hbar ck ((\tilde{\bm{\eta}}^{*}(\bm{k}) \cdot \tilde{\bm{\eta}}(\bm{k})) - k^{-2} (\tilde{\bm{\eta}}^{*}(\bm{k}) \cdot \bm{k})(\tilde{\bm{\eta}}(\bm{k}) \cdot \bm{k})), \nonumber \\ (\Delta H(\gamma))^{2} &=& \int \frac{d\bm{k}}{(2\pi)^{3}} 2\pi\hbar ck ((\tilde{\bm{\gamma}}^{*}(\bm{k}) \cdot \tilde{\bm{\gamma}}(\bm{k})) - k^{-2} (\tilde{\bm{\gamma}}^{*}(\bm{k}) \cdot \bm{k})(\tilde{\bm{\gamma}}(\bm{k}) \cdot \bm{k})).
\end{eqnarray}

Hereafter we assume a Gaussian gate, i.e., $\bm{\eta}(\bm{r}) = \bm{u} \delta (\bm{r}; l)$ and $\bm{\gamma}(\bm{r}) = \bm{v} \delta (\bm{r} - \bm{r}_{0}; l)$, with $\delta (\bm{r}; l)$ given by Eq.~(\ref{delta-Gauss}), and $\bm{u}$ and $\bm{v}$ are unit vectors. A simple inspection of Eq.~(\ref{Delta-E-H-continuous}) shows that the uncertainties do not depend on a particular choice of $\bm{u}$, $\bm{v}$, and $\bm{r}_{0}$ (reflecting the Poincare symmetry of the vacuum state), and, for our case, we have $\Delta H(\gamma) = \Delta E(\eta)$, so that we need only compute $\Delta E(\eta)$. Choosing the $z$-axis along $\bm{u}$, and applying the spherical coordinates, we obtain upon substitution of the Fourier transform
\begin{eqnarray}
\label{eta-Gauss-Fourier} \tilde{\eta}_{j}(\bm{k}) = e^{-(l^{2}k^{2}/2)}
\end{eqnarray}
of $\delta(\bm{r}; l)$ into Eq.~(\ref{Delta-E-H-continuous})
\begin{eqnarray}
\label{Delta-E-Gauss-coh} (\Delta E(\eta))^{2} &=& \frac{2\pi\hbar c}{(2\pi)^{3}} \int_{0}^{\infty}k^{2}dk \int_{0}^{2\pi}d\varphi \int_{0}^{\pi} \sinh\theta d\theta k (1 - \sin^{2}\theta) e^{-l^2 k^{2}} = \frac{\hbar c}{2\pi} \int_{-1}^{1}d\tau\tau^{2} \int_{0}^{\infty} dk k^{3} e^{-l^2 k^{2}} \nonumber \\ &=& \frac{1}{3\pi} \frac{\hbar c}{l^{4}},
\end{eqnarray}
so that
\begin{eqnarray}
\label{Delta-E-H-Gauss-coh} (\Delta E(\eta))^{2} = (\Delta H(\gamma))^{2} = \frac{1}{3\pi} \frac{\hbar c}{l^{4}}.
\end{eqnarray}

In section~\ref{sec:EM-quant-uncert-G-gate} we identified the maximal value of the r.h.s. of the Heisenberg uncertainty principle for EM-field for Gaussian gates of identical shape, with respect to the shift between the gates [see Eq.~(\ref{E-H-Heisenberg-example})], so that the uncertainty Heisenberg principle is indeed satisfied
\begin{eqnarray}
\label{Delta-E-H-Gauss-coh-2} (\Delta E(\eta)) \cdot (\Delta H(\gamma)) = \frac{1}{3\pi} \frac{\hbar c}{l^{4}} > \frac{1}{2 \sqrt{2\pi} \sqrt{e}} \frac{\hbar c}{l^{4}}.
\end{eqnarray}

The  commutator of $\hat{E}$ and $\hat{H}$ is a number which determines the lower bound of their uncertainty. This bound does not depend on the quantum state of the field and exists also in the vacuum state. We can then associate it with vacuum fluctuations.
The $\Delta E$ and $\Delta H$ uncertainties do depend on the quantum state of light. We thus found that the vacuum state (and hence the coherent state) fluctuations are larger than the minimal uncertainty. This is similar to the results, reported in~\cite{Gureyev-17} that the minimal product of precision in intensity and degree of spatial localization is achieved not at coherent states, but rather at the Epanechnikov distributions. Points of future interest include (i) finding a quantum state that satisfies the minimum uncertainty, (ii) investigating the connection of the presented uncertainty relations to the intensity vs spatial localization counterparts, explicitly considered in~\cite{Gureyev-17}, and (iii) minimizing $\Delta E$ or $\Delta H$ to create squeezed fluctuations \cite{Kuznetsov-16,Zeng-18}. Another future goal is to identify an experimental signal  that is sensitive to this uncertainty, e.g., chirality in a nanostructure \cite{Potma-19,Langer-20,Pozzi-17}.

An alternative experimental realization of the present ideas is possible using nano-structures, rather than spatial gating. Let us consider a setup whereby a nano sample interacts with a small part of a beam. The entire beam has many photons and is classical but the relevant part of the beam  that interacts with matter in this measurement is only a nano slice that contains very few photons and should therefore be described by a quantum light. The measurement should thus show quantum light effects even though it employs a classical field! A simple analogy of this state of affairs exits in the time/frequency (rather than space) domain. When a short broad-band pulse interacts with a system which has a narrow spectral line, only few of its spectral components participate, and the experiment can be described by an effective narrow band (and long) pulse. The temporal resolution is eroded and not given by the pulse duration. The experiment may be then described by an effective pulse which is different from the incoming pulse \cite{Polli-10}. This is the temporal analogue of the spatial gating discussed here.

\acknowledgments
VYC gratefully acknowledges the support of the U.S. Department of Energy, Office of Sciences, Material Sciences and Engineering Division, Condensed Matter Theory Program. SM gratefully acknowledges the support of the National Science Foundation (Grant CHE-1953045).

\section*{DATA AVAILABILITY}
Data sharing is not applicable to this article as no new data were created or analyzed in this study

\appendix


\section{Quantization of the Electromagnetic Field and the Heisenberg Uncertainty Relation}
\label{sec:EM-quant-uncert}

In this appendix we derive the commutation relation [Eq.~(\ref{expression-for-c})] between the gated electric and magnetic field operators. We start with quantization of electromagnetic field in vacuum, considered as a dynamical system. Its classical action is given by
\begin{eqnarray}
\label{EM-action} && S [\bm{A}(\bm{r}), A_{0}(\bm{r}); \dot{\bm{A}}(\bm{r}), \dot{A}_{0}(\bm{r})] = \int dt L; \;\;\; L = \frac{1}{8\pi} \int d\bm{r} (\bm{E}^{2} - \bm{H}^{2}); \nonumber \\ && \bm{E} = c^{-1} \dot{\bm{A}} - \bm{\partial} A_{0}; \;\;\; H_{i} = \varepsilon_{ijk} \partial_{j}A_{k}, \;\;\; \partial_{j} = \frac{\partial}{\partial r_{j}}.
\end{eqnarray}

To switch to the (classical) Hamilton formalism, we identify the conjugate momenta as variational derivatives
\begin{eqnarray}
\label{EM-momenta} \pi_{j}(\bm{r}) = \frac{\delta L}{\delta A_{j}(\bm{r})} = \frac{1}{4\pi c} E_{j}(\bm{r}); \;\;\; \pi_{0}(\bm{r}) = \frac{\delta L}{\delta A_{0}(\bm{r})} = \frac{1}{4\pi c} E_{0}(\bm{r}) = 0,
\end{eqnarray}
the second relation $\pi_{0}(\bm{r}) = 0$ indicates that we are dealing with a dynamical system with constraints. The classical Hamiltonian is obtained in a standard way
\begin{eqnarray}
\label{EM-H-classical} {\cal H} &=& \int d\bm{r} (\pi_{0}(\bm{r}) \dot{A}_{0}(\bm{r}) + \bm{\pi}(\bm{r}) \cdot \dot{\bm{A}}(\bm{r})) - L = \frac{1}{8\pi} \int d\bm{r} (\bm{E}^{2} + \bm{H}^{2}) + \frac{1}{4\pi} \int d\bm{r} \bm{E} \cdot \bm{\partial} A_{0} \nonumber \\ &=& \frac{1}{8\pi} \int d\bm{r} (\bm{E}^{2} + \bm{H}^{2}) - \frac{1}{4\pi} \int d\bm{r} A_{0} {\rm div} \, \bm{E},
\end{eqnarray}
where we have made use of Eq.~(\ref{EM-momenta}). The Poisson bracket has a standard form
\begin{eqnarray}
\label{EM-P-bracket} \{E_{j}(\bm{r}), A_{k}(\bm{r}')\} = 4\pi c \delta_{jk} \delta(\bm{r} - \bm{r}'), \;\;\; \{E_{0}(\bm{r}), A_{0}(\bm{r}')\} = 4\pi c \delta(\bm{r} - \bm{r}').
\end{eqnarray}
For the constraint $E_{0}(\bm{r})$ to be preserved by the dynamics, we need to have $\dot{E}_{0}(\bm{r}) = 0$, which combined with the Hamilton equation $\dot{E}_{0}(\bm{r}) = \{{\cal H}, \dot{E}_{0}(\bm{r})\}$ and upon a direct computation of the r.h.s. of the latter [making use of Eqs.~(\ref{EM-H-classical}) and (\ref{EM-P-bracket})], yields ${\rm div} \, \bm{E}(\bm{r}) = 0$, referred to as the secondary constraints. Computation of the time derivative of the secondary constraints yields $0$, meaning that there are no higher-order constraints, so that the complete set of them is given by
\begin{eqnarray}
\label{EM-constraints} E_{0}(\bm{r}) = 0, \;\;\; {\rm div} \, \bm{E}(\bm{r}) = 0;
\end{eqnarray}
obviously the Poisson bracket between constraint is zero, so that all constraints are type-1 in Dirac's classification.

Gauge invariant quantization is performed using a canonical quantization of a dynamical system with type-2 (weak) constraints. A state in the extended Hilbert space is represented by a wavefunction $\Psi [\bm{A}(\bm{r}), A_{0}(\bm{r})]$ with the electric field operators naturally defined as variational derivatives
\begin{eqnarray}
\label{EM-E-operators} \hat{E}_{0}(\bm{r}) = - 4\pi ci\hbar \frac{\delta}{\delta A_{0}(\bm{r})}, \;\;\; \hat{E}_{j}(\bm{r}) = - 4\pi ci\hbar \frac{\delta}{\delta A_{j}(\bm{r})}.
\end{eqnarray}
The constraints are applied in a weak sense, i.e., we introduce a physical subspace of the wavefunctions that satisfy the conditions $\hat{E}_{0}(\bm{r}) \Psi = 0$ and ${\rm div} \,\hat{\bm{E}}(\bm{r}) \Psi = 0$, or explicitly
\begin{eqnarray}
\label{EM-phys-Hilbert} \frac{\delta}{\delta A_{0}(\bm{r})} \Psi = 0, \;\;\; \frac{\partial}{\partial r_{j}} \frac{\delta}{\delta A_{j}(\bm{r})} \Psi = 0,
\end{eqnarray}
which means that a physical wavefunction is a one that does not depend on the scalar potential and the longitudinal component of the vector counterpart. A physical, or equivalently a gauge invariant, operator is a one that acts closely in the physical subspace.

To simplify the application of constraints we switch to the momentum domain. To that end we consider the space as a box of size $L$ (with volume $V$) with periodic boundary conditions, so that the quantization conditions for the photon wavevector are $\bm{k} L = 2\pi \bm{n}$, with $\bm{n} \in \mathbb{Z}^{3}$, allowing to represent
\begin{eqnarray}
\label{coord-to-momentum} \bm{A}(\bm{r}) = \frac{1}{\sqrt{V}} \sum_{\bm{k}} \bm{A}_{\bm{k}} e^{i \bm{k} \cdot \bm{r}}, \;\;\; \bm{A}_{\bm{k}} = \frac{1}{\sqrt{V}} \int_{X} d\bm{r} \bm{A}(\bm{r}) e^{-i \bm{k} \cdot \bm{r}}.
\end{eqnarray}
We further associate with each allowed $\bm{k}$ an orthonormal basis set $(\bm{e}_{\bm{k}}^{(\alpha)} \, | \, \alpha = 1, 2, 3)$, with $\bm{e}_{\bm{k}}^{(\alpha)} = \bm{n} = |\bm{k}|^{-1} \bm{k}$ and represent
\begin{eqnarray}
\label{coord-to-momentum-2} \bm{A}_{\bm{k}} = \sum_{\alpha = 1}^{3}A_{\alpha, \bm{k}} \bm{e}_{\bm{k}}^{(\alpha)}, \;\;\; \bm{E }_{\bm{k}} = \sum_{\alpha = 1}^{3} E_{\alpha, \bm{k}} \bm{e}_{\bm{k}}^{(\alpha)},
\end{eqnarray}
resulting in
\begin{eqnarray}
\label{EM-P-bracket-k} \{E_{\alpha, \bm{k}} , A_{\beta, \bm{k}'}\} &=& \frac{4\pi c}{V} \int d\bm{r} d\bm{r}' e^{i \bm{k} \cdot \bm{r} + i \bm{k}' \cdot \bm{r}'} \{\bm{e}_{\bm{k}}^{(\alpha)} \cdot \bm{E}(\bm{r}), \bm{e}_{\bm{k}'}^{(\beta)} \cdot \bm{A}(\bm{r}')\} \nonumber \\ &=& 4\pi c \int \frac{d\bm{r} d\bm{r}'}{V} e^{i \bm{k} \cdot \bm{r} + i \bm{k}' \cdot \bm{r}'} \bm{e}_{\bm{k}}^{(\alpha)} \cdot \bm{e}_{\bm{k}'}^{(\beta)} \delta (\bm{r} - \bm{r}') = 4\pi c \bm{e}_{\bm{k}}^{(\alpha)} \cdot \bm{e}_{\bm{k}'}^{(\beta)} \delta_{\bm{k}+\bm{k}', 0}.
\end{eqnarray}

Applying the constraints we have $E_{\bm{k}}^{(3)} = 0$, so that
\begin{eqnarray}
\label{EM-P-bracket-k-2} \bm{E}(\bm{r}) = \frac{1}{\sqrt{V}} \sum_{\bm{k}} \sum_{\alpha=1}^{2} \bm{e}_{\bm{k}}^{(\alpha)} e^{i \bm{k} \cdot \bm{r}} E_{\alpha, \bm{k}}, \;\;\; \bm{H}(\bm{r}') = \frac{i}{\sqrt{V}} \sum_{\bm{k}'} \sum_{\beta=1}^{2} [\bm{k}', \bm{e}_{\bm{k}'}^{(\beta)}] e^{i \bm{k}' \cdot \bm{r}'} A_{\beta, \bm{k}'},
\end{eqnarray}
and denoting with $(\bm{u}_{j} \, | \, j = 1, 2, 3)$ a lab frame, we compute
\begin{eqnarray}
\label{EM-P-bracket-k-3} \{E_{j}(\bm{r}), H_{s}(\bm{r}')\} &=& \{\bm{u}_{j} \cdot \bm{E}(\bm{r}), \bm{u}_{s} \cdot \bm{H}(\bm{r}')\} = \frac{i}{V} \sum_{\bm{k} \bm{k}'} \sum_{\alpha\beta} (\bm{e}_{\bm{k}}^{(\alpha)} \cdot \bm{u}_{j})([\bm{k}',\bm{e}_{\bm{k}'}^{(\beta)}] \cdot \bm{u}_{s}) e^{i \bm{k} \cdot \bm{r} + i \bm{k}' \cdot \bm{r}'} \{E_{\alpha, \bm{k}}, A_{\beta, \bm{k}'}\} \nonumber \\ &=& - \frac{4\pi c i}{V} \sum_{\bm{k}} \sum_{\alpha\beta} (\bm{e}_{\bm{k}}^{(\alpha)} \cdot \bm{u}_{j})([\bm{k},\bm{e}_{-\bm{k}}^{(\beta)}] \cdot \bm{u}_{s}) (\bm{e}_{\bm{k}}^{(\alpha)} \cdot \bm{e}_{-\bm{k}}^{(\beta)}) e^{i \bm{k} \cdot (\bm{r} - \bm{r}')},
\end{eqnarray}
where the summation in Eq.~(\ref{EM-P-bracket-k-3}) runs over $\alpha, \beta = 1, 2$. We further compute
\begin{eqnarray}
\label{EM-P-bracket-k-4} && \sum_{\alpha\beta} (\bm{e}_{\bm{k}}^{(\alpha)} \cdot \bm{u}_{j})([\bm{k},\bm{e}_{-\bm{k}}^{(\beta)}] \cdot \bm{u}_{s}) (\bm{e}_{\bm{k}}^{(\alpha)} \cdot \bm{e}_{-\bm{k}}^{(\beta)}) = \sum_{\alpha\beta} (\bm{e}_{\bm{k}}^{(\alpha)} \cdot \bm{u}_{j})([\bm{u}_{s}, \bm{k}] \cdot \bm{e}_{-\bm{k}}^{(\beta)}) (\bm{e}_{\bm{k}}^{(\alpha)} \cdot \bm{e}_{-\bm{k}}^{(\beta)}) \nonumber \\ && \;\;\; = (\sum_{\alpha} (\bm{u}_{j} \cdot \bm{e}_{\bm{k}}^{(\alpha)}) \bm{e}_{\bm{k}}^{(\alpha)}) \cdot (\sum_{\beta} ([\bm{u}_{s}, \bm{k}] \cdot \bm{e}_{-\bm{k}}^{(\beta)}) \bm{e}_{-\bm{k}}^{(\beta)}) = (\bm{u}_{j} \cdot [\bm{u}_{s}, \bm{k}]) = [\bm{u}_{j}, \bm{u}_{s}] \cdot \bm{k} = \varepsilon_{jsm} (\bm{u}_{m} \cdot \bm{k}) = \varepsilon_{jsm} k_{m}.
\end{eqnarray}
Note that in the first expression in the second line in Eq.~(\ref{EM-P-bracket-k-4}) the summation runs over $\alpha, \beta = 1, 2$, whereas to obtain the next equality in the chain, the summation has to be extended to its full range $\alpha, \beta = 1, 2, 3$. This is indeed allowed, since the summand with $\beta = 3$ turns out to $0$, which also implies that the scalar product of the summand with $\alpha = 3$ with the sum over $\beta$ also turns out to zero, so that the summations can be safely extended to their full ranges. Upon the substitution of Eq.~(\ref{EM-P-bracket-k-4}) into Eq.~(\ref{EM-P-bracket-k-3}), we obtain
\begin{eqnarray}
\label{EM-P-bracket-k-5} \{E_{j}(\bm{r}), H_{s}(\bm{r}')\} = - \frac{4\pi c}{V} \sum_{\bm{k}} \varepsilon_{jsm} ik_{m} e^{i \bm{k} \cdot (\bm{r} - \bm{r}')} = - 4\pi c \varepsilon_{jsm} \frac{\partial}{\partial r_{m}} \frac{1}{V} \sum_{\bm{k}} e^{i \bm{k} \cdot (\bm{r} - \bm{r}')} = - 4\pi c \varepsilon_{jsm} \frac{\partial}{\partial r_{m}} \delta (\bm{r} - \bm{r}').
\end{eqnarray}

Finally, we make use of Eq.~(\ref{EM-P-bracket-k-5}) to compute the Poisson bracket of the gated variables
\begin{eqnarray}
\label{EM-P-bracket-k-6} \{E(\eta), H(\gamma)\} &=& - 4\pi c \varepsilon_{jsm} \int d\bm{r} d\bm{r}' \eta_{j}(\bm{r}) \gamma_{s}(\bm{r}') \frac{\partial}{\partial r_{m}} \delta(\bm{r} - \bm{r}') = 4\pi c \varepsilon_{jsm} \int d\bm{r} d\bm{r}' \delta(\bm{r} - \bm{r}') \frac{\partial \eta_{j}(\bm{r})}{\partial r_{m}} \gamma_{s}(\bm{r}) \nonumber \\ &= & 4\pi c \int d\bm{r} \varepsilon_{smj} \frac{\partial \eta_{j}(\bm{r})}{\partial r_{m}} \gamma_{s}(\bm{r}) = 4\pi c \int d\bm{r} ({\rm curl} \, \bm{\eta}(\bm{r})) \cdot \bm{\gamma}(\bm{r}) = h (\eta, \gamma).
\end{eqnarray}
with $h (\eta, \gamma)$ given by Eq.~(\ref{expression-for-c}).

Upon canonical gauge invariant quantization of the electromagnetic field, Eq.~(\ref{EM-P-bracket-k-6}) provides the expression for the commutator of the gated operators, given by Eq.~(\ref{commute-uncertain}).

\section{Uncertainties of the Gated Electric and Magnetic Fields in a Coherent State: Details}
\label{sec:EM-uncertain-coherent-details}

We start with representing the EM field operators in terms of the creation/annihilation operators
\begin{eqnarray}
\label{E-A-to-a} \hat{A}_{\alpha, \bm{k}} = u_{\bm{k}} a_{\alpha, \bm{k}} + u_{\bm{k}}^{*} a_{\alpha, -\bm{k}}^{\dagger}, \;\;\; \hat{E}_{\alpha, \bm{k}} = v_{\bm{k}} a_{\alpha, \bm{k}} + v_{\bm{k}}^{*} a_{\alpha, -\bm{k}}^{\dagger},
\end{eqnarray}
together with a choice $\bm{e}_{-\bm{k}}^{(\alpha)} = \bm{e}_{\bm{k}}^{(\alpha)}$, $u_{-\bm{k}} = u_{\bm{k}}$, $u_{-\bm{k}} = u_{\bm{k}}$  so that the conditions $\hat{A}_{\alpha, -\bm{k}} = \hat{A}_{\alpha, \bm{k}}^{\dagger}$ and $\hat{E}_{\alpha, -\bm{k}} = \hat{E}_{\alpha, -\bm{k}}^{\dagger}$ that reflect the real nature of the electromagnetic field are satisfied. Upon quantizing the expression for the Poisson bracket [Eq.~\ref{EM-P-bracket-k})] and postulating the commutation relations for the photon operators we obtain
\begin{eqnarray}
\label{EM-commute-k} [\hat{E}_{\alpha, \bm{k}}, \hat{A}_{\beta, \bm{k}'}] = -4\pi i\hbar c \bm{e}_{\bm{k}}^{(\alpha)} \cdot \bm{e}_{\bm{k}'}^{(\beta)} \delta_{\bm{k}+\bm{k}', 0}, \;\;\; [a_{\alpha, \bm{k}}, a_{\beta, \bm{k}}^{\dagger}] = \delta_{\alpha\beta} \delta_{\bm{k}, \bm{k}'},
\end{eqnarray}
by substituting Eq.~(\ref{E-A-to-a}) into Eq.~(\ref{EM-commute-k}), we obtain
\begin{eqnarray}
\label{EM-commute-k-2}  -4\pi i \hbar c \delta_{\alpha\alpha'} \delta_{\bm{k}, -\bm{k}'} &=& -4\pi i \hbar c \bm{e}_{\bm{k}}^{(\alpha)} \cdot \bm{e}_{\bm{k}'}^{(\alpha')} \delta_{\bm{k}+\bm{k}', 0} = [v_{\bm{k}} a_{\alpha, \bm{k}} + v_{\bm{k}}^{*} a_{\alpha, -\bm{k}}^{\dagger}, u_{\bm{k}'} a_{\alpha', \bm{k}'} + u_{\bm{k}'}^{*} a_{\alpha', -\bm{k}'}^{\dagger}] \nonumber \\ &=& (v_{\bm{k}} u_{\bm{k}}^{*} - v_{\bm{k}}^{*} u_{\bm{k}}) \delta_{\alpha\alpha'} \delta_{\bm{k}, -\bm{k}'}, \;\; {\rm implying} \;\; v_{\bm{k}} u_{\bm{k}}^{*} - v_{\bm{k}}^{*} u_{\bm{k}} = -i 4\pi \hbar c.
\end{eqnarray}

We further have
\begin{eqnarray}
\label{EM-H} \hat{E}_{\alpha, \bm{k}} \hat{E}_{\alpha, -\bm{k}} &=& v_{\bm{k}}^{*} v_{\bm{k}} (a_{\alpha, \bm{k}}^{\dagger} a_{\alpha, \bm{k}} + a_{\alpha, -\bm{k}}^{\dagger} a_{\alpha, -\bm{k}}) + v_{\bm{k}}^{*} v_{\bm{k}} + (v_{\bm{k}}^{2} a_{\alpha, \bm{k}} a_{\alpha, \bm{k}} + (v_{\bm{k}}^{*})^{2} a_{\alpha, -\bm{k}}^{\dagger} a_{\alpha, \bm{k}}^{\dagger}), \nonumber \\ \hat{A}_{\alpha, \bm{k}} \hat{A}_{\alpha, -\bm{k}} &=& u_{\bm{k}}^{*} u_{\bm{k}} (a_{\alpha, \bm{k}}^{\dagger} a_{\alpha, \bm{k}} + a_{\alpha, -\bm{k}}^{\dagger} a_{\alpha, -\bm{k}}) + u_{\bm{k}}^{*} u_{\bm{k}} + (u_{\bm{k}}^{2} a_{\alpha, \bm{k}} a_{\alpha, \bm{k}} + (u_{\bm{k}}^{*})^{2} a_{\alpha, -\bm{k}}^{\dagger} a_{\alpha, \bm{k}}^{\dagger}),
\end{eqnarray}
the Hamiltonian ${\cal H}$ of the EM-field in vacuum reads
\begin{eqnarray}
\label{EM-H-2} {\cal H} = \sum_{\bm{k}} \sum_{\alpha} {\cal H}_{\alpha, \bm{k}},
\end{eqnarray}
with
\begin{eqnarray}
\label{EM-H-3} {\cal H}_{\alpha, \bm{k}} &=& \frac{1}{8\pi} (v_{\bm{k}}^{*} v_{\bm{k}} + k^{2} u_{\bm{k}}^{*} u_{\bm{k}}) (a_{\alpha, \bm{k}}^{\dagger} a_{\alpha, \bm{k}} + a_{\alpha, -\bm{k}}^{\dagger} a_{\alpha, -\bm{k}}) + \frac{1}{8\pi} (v_{\bm{k}}^{*} v_{\bm{k}} + k^{2} u_{\bm{k}}^{*} u_{\bm{k}}) \nonumber \\ &+& \frac{1}{8\pi} (v_{\bm{k}}^{2} + k^{2}u_{\bm{k}}^{2}) a_{\alpha, \bm{k}} a_{\alpha, -\bm{k}} + \frac{1}{8\pi} (v_{\bm{k}}^{2} + k^{2}u_{\bm{k}}^{2})^{*} a_{\alpha, \bm{k}}^{\dagger} a_{\alpha, -\bm{k}}^{\dagger}.
\end{eqnarray}
Combining the condition that the unwanted terms [the last two terms in Eq.~(\ref{EM-H-3})] disappear with Eq.~(\ref{EM-commute-k-2}) we obtain a system of two equations
\begin{eqnarray}
\label{EM-H-4} v_{\bm{k}}^{2} + k^{2}u_{\bm{k}}^{2} = 0, \;\;\; v_{\bm{k}} u_{\bm{k}}^{*} - v_{\bm{k}}^{*} u_{\bm{k}} = -i 4\pi \hbar c
\end{eqnarray}
whose solution
\begin{eqnarray}
\label{EM-H-5} u_{\bm{k}} = \sqrt{\frac{2\pi\hbar c}{k}}, \;\;\; v_{\bm{k}} = -i \sqrt{2\pi\hbar ck},
\end{eqnarray}
results in the explicit expressions
\begin{eqnarray}
\label{EM-H-6} A_{\alpha, \bm{k}} = \sqrt{\frac{2\pi\hbar c}{k}} (a_{\alpha, \bm{k}} + a_{\alpha, -\bm{k}}^{\dagger}), \;\;\; E_{\alpha, \bm{k}} = -i \sqrt{2\pi\hbar ck} (a_{\alpha, \bm{k}} - a_{\alpha, -\bm{k}}^{\dagger}).
\end{eqnarray}

Upon substitution of Eq.~(\ref{EM-H-5}) into Eqs.~(\ref{EM-H-2}) and (\ref{EM-H-3}) we obtain the standard expression for the Hamiltonian ${\cal H}$ of the EM-field, recast in terms of the photon creation/annihilation operators
\begin{eqnarray}
\label{EM-H-7} {\cal H} = {\cal E}_{0} + \sum_{\bm{k}} \sum_{\alpha} \hbar ck a_{\alpha, \bm{k}}^{\dagger} a_{\alpha, \bm{k}}, \;\;\; {\cal E}_{0} = \frac{1}{2} \sum_{\bm{k}} \sum_{\alpha} \hbar ck.
\end{eqnarray}

Upon the substitution of Eqs.~(\ref{EM-P-bracket-k-2}) and (\ref{EM-H-6}) into Eq.~(\ref{E-H-gated-operators}) we obtain
\begin{eqnarray}
\label{compute-Delta-E} \hat{E}(\eta) = \int d{\bm{r}} \bm{\eta}(\bm{r}) \cdot \bm{E}(\bm{r}) = \frac{1}{\sqrt{V}} \sum_{\bm{k}\alpha} \int d{\bm{r}} \bm{\eta}(\bm{r}) \cdot \bm{e}_{\bm{k}}^{(\alpha)} E_{\alpha, \bm{k}} = -i \sum_{\bm{k}\alpha} (\bm{\eta}_{-\bm{k}} \cdot \bm{e}_{\bm{k}}^{(\alpha)}) i \sqrt{2\pi\hbar ck} (a_{\alpha, \bm{k}} - a_{\alpha, -\bm{k}}^{\dagger}),
\end{eqnarray}
and further, making use of Eq.~(\ref{compute-Delta-E}) and applying abbreviated notation $\langle \bullet \rangle = \langle \Omega|\bullet|\Omega \rangle$ we obtain
\begin{eqnarray}
\label{compute-Delta-E-2} (\Delta E(\eta))^{2} &=& \langle (\hat{E}(\eta))^{2} \rangle = - \sum_{\bm{k}\alpha} \sum_{\bm{k}'\alpha'} (\bm{\eta}_{-\bm{k}} \cdot \bm{e}_{\bm{k}}^{(\alpha)}) (\bm{\eta}_{-\bm{k}'} \cdot \bm{e}_{\bm{k}'}^{(\alpha')}) \sqrt{2\pi\hbar ck} \sqrt{2\pi\hbar ck'} \langle(a_{\alpha, \bm{k}} - a_{\alpha, -\bm{k}}^{\dagger}) (a_{\alpha', \bm{k}'} - a_{\alpha', -\bm{k}'}^{\dagger})\rangle \nonumber \\ &=& \sum_{\bm{k}\alpha} \sum_{\bm{k}'\alpha'} (\bm{\eta}_{-\bm{k}} \cdot \bm{e}_{\bm{k}}^{(\alpha)}) (\bm{\eta}_{-\bm{k}'} \cdot \bm{e}_{\bm{k}'}^{(\alpha')}) \sqrt{2\pi\hbar ck} \sqrt{2\pi\hbar ck'} \delta_{\bm{k}, -\bm{k}'} \delta_{\alpha\alpha'} = \sum_{\bm{k}\alpha} 2\pi\hbar ck (\bm{\eta}_{-\bm{k}} \cdot \bm{e}_{\bm{k}}^{(\alpha)}) (\bm{\eta}_{\bm{k}} \cdot \bm{e}_{\bm{k}}^{(\alpha)}) \nonumber \\ &=& \sum_{\bm{k}} 2\pi\hbar ck (\bm{\eta}_{-\bm{k}} - k^{-2}(\bm{\eta}_{-\bm{k}} \cdot \bm{k}) \bm{k}) \cdot (\bm{\eta}_{\bm{k}} - k^{-2}(\bm{\eta}_{\bm{k}} \cdot \bm{k}) \bm{k}) \nonumber \\ &=& \sum_{\bm{k}} 2\pi\hbar ck ((\bm{\eta}_{-\bm{k}} \cdot \bm{\eta}_{\bm{k}}) - k^{-2} (\bm{\eta}_{-\bm{k}} \cdot \bm{k})(\bm{\eta}_{\bm{k}} \cdot \bm{k})) = \sum_{\bm{k}} 2\pi\hbar ck ((\bm{\eta}_{\bm{k}}^{*} \cdot \bm{\eta}_{\bm{k}}) - k^{-2} (\bm{\eta}_{\bm{k}}^{*} \cdot \bm{k})(\bm{\eta}_{\bm{k}} \cdot \bm{k})).
\end{eqnarray}
A similar computation for the uncertainty of the magnetic field yields
\begin{eqnarray}
\label{Delta-H} (\Delta H(\gamma))^{2} = \sum_{\bm{k}} 2\pi\hbar ck ((\bm{\gamma}_{\bm{k}}^{*} \cdot \bm{\gamma}_{\bm{k}}) - k^{-2} (\bm{\gamma}_{\bm{k}}^{*} \cdot \bm{k})(\bm{\gamma}_{\bm{k}} \cdot \bm{k})).
\end{eqnarray}
These result in Eq.~(\ref{Delta-E-H-continuous}) in the main text.

\end{document}